\documentstyle[preprint,cite,aps]{revtex}
\begin{document}
\draft
\title{Radiative Corrections to the Aharonov-Bohm Scattering}
\author{L.C. de Albuquerque, M. Gomes
and  A.J. da Silva\footnote{e-mails:{\it ajsilva,claudio,
mgomes@fma.if.usp.br}}}
\address{Instituto de F\'\i sica, USP\\
 C.P. 66318 - 05389-970, S\~ao\ Paulo - SP,  Brazil}
\date{May 1999}
\maketitle
\begin{abstract}
  We consider the scattering of relativistic electrons from a thin
  magnetic flux tube and perturbatively calculate the order $\alpha$,
  radiative correction, to the first order Born approximation. We show
  also that the second order Born amplitude vanishes, and obtain a
  finite inclusive cross section for the one-body scattering which
  incorporates soft photon bremsstrahlung effects. Moreover, we
  determine the radiatively corrected Aharonov-Bohm potential and, in
  particular, verify that an induced magnetic field is generated
  outside of the flux tube.
\end{abstract}
 
\vspace{4cm}

\vfill
\newpage

\section{Introduction}

Much effort has been devoted in recent years to fully understand 
the consequences as well as the mathematical subtleties of the 
Aharonov-Bohm (AB) effect \cite{AB}. The motivations range from 
different areas as low dimensional condensed matter physics 
(e.g. in the study of anyons) \cite{Anyons} to cosmic string models 
\cite{Strings} and have both experimental \cite{Tono}
and theoretical overtones.  The conceptual aspects are not less
interesting as we briefly illustrate.  Basically due to the use of
different boundary conditions, to achieve accordance between the exact
and the perturbative calculations in the case of spin zero particles,
it was necessary to include in the perturbative method a contact delta
like interaction. In 2+1 dimensions, within the Chern-Simons \cite{Jac} 
field theory approach,
it was shown that the contact interaction may be simulated by a
quartic self-interaction of the matter field, with a coupling tuned to
eliminate divergences and restore the conformal invariance of the tree
approximation \cite{Lozano}. For the scattering of two spin up fermions
it was verified \cite{Hagen0} that an additional self-interaction was
not needed since its role was provided by Pauli's magnetic term.
However, if the fermions had anti-parallel spins the effect of the
magnetic interaction canceled and a divergence showed up. These
problems were studied in a relativistic quantum field theory approach
\cite{MG1,Gomes1}.  Differently from the non-relativistic calculations
previously mentioned, without any additional hypothesis, the
scattering amplitudes are finite for both parallel and
anti-parallel spin fermions \cite{Gomes1}.

In this paper we extend these investigations by considering the AB scattering
of spin 1/2 particles by an external flux tube,
directly in 3+1 dimensions. We calculate the elastic cross 
section, in first order Born approximation in the AB potential,
including radiative corrections
up to order $\alpha$. The second order Born approximation 
is also calculated and found to vanish.
Infrared divergences of the elastic cross section, 
are eliminated from the observable scattering process by considering 
the \lq\lq inclusive'' cross section, which besides the elastic process,
also incorporates soft photon bremsstrahlung contributions. 
The discussion of vacuum polarization effects allows us to
correct the original AB potential, showing a screening of the magnetic 
flux and the induction of a magnetic field outside of the original 
flux tube. 

The paper is organized as follows. In Sec. \ref{SA} we introduce the
model, calculate the first order Born approximation including radiative 
corrections up to order $\alpha$, calculate
the second order Born approximation, and
the finite inclusive cross section including soft photon bremsstrahlung.
In Sec. \ref{SB} we obtain, up to the same order, the radiatively
corrected  AB potential, and the induced magnetic field outside the flux 
tube.   
Some comments are made in the Conclusions. 

\section{Radiative Corrections to the Scattering Cross Section}
\label{SA}

Our starting point is the standard Feynman gauge QED 
Lagrangian \cite{Zuber}
\begin{eqnarray}
\label{1}
& &{\cal L}=\bar\psi[i\gamma^\mu(\partial_\mu +iea_\mu+ieA_\mu)-m]\psi
-\frac{1}{4}f_{\mu\nu}f^{\mu\nu}-\frac{1}{2}(\partial_\mu a^\mu)^2,\\
& & f_{\mu\nu}=\partial_\mu a_\nu-\partial_\nu a_\mu.\nonumber
\end{eqnarray}
where $a_\mu(x)$ is the radiation field and $A_\mu(x)$ is
the external AB potential ($e=-|e|$ is the electron
charge). The magnetic flux tube is located at the $z$-axis, and
the AB potential can be chosen as (the index $t$ stands for 
\lq\lq transverse'' to the $z$-axis)

\begin{eqnarray}
\label{2}
& &A^\mu(x)=(0,\vec{A}_t(\vec{x}_t),0),\nonumber\\
& &A^i(\vec{x}_t)=-\frac{\Phi}{2\pi}\epsilon^{ij}\frac{x^j}{\rho^2},\;\;\;
\rho=|\vec{x}_t|=\sqrt{x_1^2+x_2^2}\neq 0.
\end{eqnarray}
where $\Phi$ is the magnetic flux\footnote{\footnotesize We use 
natural units $(c=\hbar=1)$, and
diag$\,g_{\mu\nu}=(+,-,-,-)$. Repeated Greek indices sum from 0 to 3, 
while  repeated Latin indices from the beginning and from the middle of the 
alphabet sum from 1 to 3 and from 1 to 2, respectively. The antisymmetric
tensor $\epsilon^{ij}$ is normalized such that $\epsilon^{12}=1$.}.
The  AB potential satisfies the Coulomb gauge condition,
$\partial^i A_i=0$, and in cylindrical polar coordinates it reads
$A_\varphi(\rho)=\frac{\Phi}{2\pi \rho}$.

Starting from the Lagrangian (\ref{1}), there are in principle two ways
to proceed: one could try to construct the exact Feynman propagator in
the background potential (\ref{2}), and then evaluate the radiative
corrections to the scattering amplitude, or one can resort to
a perturbative calculation both in $\beta=\frac{e\Phi}{2 \pi}$ and
$\alpha=\frac{e^2}{4\pi}$.  In the present paper we will restrict
ourselves to the second approach.
Specifically, we have to calculate the first order Born amplitude in
$\beta$ plus radiative corrections up to order $\alpha$, hereafter
designated by $T_1$ and add these contributions to the second order
Born amplitude in $\beta$, designated by $T_2$ (see figure {\bf 1}):

\begin{eqnarray}
& &-i T = -iT_1 -iT_2,\nonumber\\
\label{3}
& &-iT_1=(-ie)\bar{u}(p^\prime,s^\prime)\Bigr\{\gamma_\rho+\Gamma_\rho(q)
+{\Pi}_{\rho\nu}(q^2)\,G^{\nu\sigma}(q^2)\,\gamma_\sigma\Bigr\}
A^{\rho}(q)\,u(p,s),\\
\label{4}
& &-iT_2=-ie^2\bar{u}(p^\prime,s^\prime)\int \frac{d^4 k}{(2\pi)^4}
\not\!\! A(p^\prime-k)\,S_F(k)\, \not\!\! A(k-p)\,u(p,s).
\end{eqnarray}
Since the scattering process  conserves  the energy
and the z-component of the three-momentum we will consider
the case of incidence perpendicular to the solenoid. Then,
in the above expressions $p^\mu=(E,\vec{p}_t,0),\,s$ and 
$p^{\prime\mu}=(E,\vec {p}^{\phantom a \prime}_t,0),
\,s^\prime$ are the four-momenta and
helicities of the incident and outgoing electrons, respectively, with
$|\vec{p}|=|\vec{p}^{\phantom a \prime}|=p_t$; 
$q^\mu=(p^\prime-p)^\mu=(0,\vec{q}_t,0)$ is the momentum transfer 
from the flux tube.  We choose a reference frame in which 
$\vec{p}_t=p_t\,(\cos\,\frac{\psi}{2},-\sin\,\frac{\psi}{2})$,
$\vec {p}^{\phantom a \prime}_t
=p_t\,(\cos\,\frac{\psi}{2},\sin\,\frac{\psi}{2})$, and
$\vec{q}_t=(0,2p_t\,\sin\,\frac{\psi}{2})$. Thus
$\psi$ is the scattering angle ($0<\psi\leq 2\pi$).

In Eqs. (\ref{3}) and (\ref{4}) $\Gamma_\rho(q)$ is the 
renormalized one-loop vertex function, to be defined later 
(see Eq. (\ref{5a})),

\begin{equation}
\label{5}
{\Pi}_{\rho\nu}(q)= -i \,\biggl(\,q^2\,g_{\rho\nu}- q_\rho q_\nu
\biggr)\,\Pi(q^2)
\end{equation}
is the renormalized vacuum polarization
tensor, $S_F(k)$ is the  fermion propagator, and
$G_{\nu\sigma}(q^2)=-i\,g_{\nu\sigma}\,\bigl(q^2-\mu^2+i\epsilon\bigr)^{-1}$
is the photon  propagator, with an infrared cutoff $\mu$.
$A_\mu(q)$ is the Fourier transform of the potential (\ref{2}) whose
non-vanishing components are
\begin{eqnarray}
\label{6}  
& &A^i(q)=(2\pi)^2\delta(q^0)\delta(q^3)\tilde{A}^i(\vec{q}_t),\nonumber\\
& &\tilde{A}^i(\vec{q}_t)=i\Phi\epsilon^{ij}\frac{q^j}{\vec{q}_t^{\phantom
a 2}+\eta^2}.
\end{eqnarray}

In Eq.(\ref{6}) we have introduced an
infrared cut-off  cut-off $\eta$ to be made zero after the calculations.
Integrating over $k_0$ and $k_3$ in Eq. (\ref{4}), and using in the 
numerator of  $S_F(k)$

\begin{equation}
\label{B1}
\not\! k+m=2m\sum_r\,u(k,r)\bar{u}(k,r),
\end{equation}
we obtain 

\begin{eqnarray}
\label{B2}
& &iT_2 =-i8\pi^2\beta^2\,m\delta(p^\prime_0-p_0)\,\delta(p^\prime_3-p_3)
\,\int d^2 k\,
\frac{\sum_r\,\omega(p^\prime\,
s^\prime;k\,r)\,\omega(k\,r;p\,s)}
{\vec{p}_t^{\phantom a 2}-\vec{k}_t^{\phantom a 2}+i\epsilon}\nonumber\\
& &\qquad\quad\equiv i(2\pi)^2\delta(p^\prime_0-p_0)\,\delta(p^\prime_3-p_3)
\tilde{T}_2,\\
\label{B3}
& &\quad\omega(p^\prime\,s^\prime;k\,r)=\bar{u}(p^\prime,s^\prime)
\,\epsilon^{ij}\,\gamma_{i}\,\frac{(p^\prime -k)_j}
{(\vec{p}^{\phantom a \prime}_t-\vec{k}_t)^{2}+\eta^2}\,u(k,r),
\end{eqnarray}
with an analogous formula for $\omega(k\,r;p\,s)$. 

The helicity basis for positive energy solutions of 
the free Dirac equation is

\begin{equation}
\label{A1}
u(\vec{k},r)=\sqrt\frac{ E+m}{2m}\,
\left(
\begin{array}{c}
\chi(r) \\
\frac{r\,k_t}{E+m}\chi(r)\\
\end{array}
\right),
\end{equation}
where $k^\mu=(E, \vec{k}_t,0)$, $k_t=|\vec{k}|$ and $r=\pm$. The
Pauli spinor $\chi(r)$ can be parametrized in spherical
polar coordinates as 

\begin{equation}
\label{A2}
\chi(r)= \frac{\sqrt{2}}{2}\,
\left(
\begin{array}{c}
1 \\
e^{i\phi}\\
\end{array}
\right)\,\delta_{r(+)}\,+\, \frac{\sqrt{2}}{2}\,
\left(
\begin{array}{c}
-e^{-i\phi}\\
1\\
\end{array}
\right)\,\delta_{r(-)},
\end{equation}
where $\phi$ is the azimuthal angle
of $\vec{k}_t$. The normalizations are $\chi^\dagger(r)\,\chi(r^\prime)
=\delta_{rr^\prime}$ and $\bar{u}(\vec{k},r)\,
u(\vec{k},r^\prime)=\delta_{rr^\prime}$.
The helicity basis for $u(\vec{p},s)$ and $u(\vec{p}^{\phantom a
\prime},s^\prime)$ are constructed in a similar way.

Using Eq. (\ref{A1}) and the similar formula for $u(\vec{p}^{\phantom a
\prime},s^\prime)$ , $\omega(p^\prime\,s^\prime;k\,r)$ is given by

\begin{eqnarray}
\label{B4}
\omega(p^\prime\,s^\prime;k\,r)=& &\frac{1}{4m}
\frac{1}{(\vec{p}_t^{\phantom a \prime}-\vec{k}_t)^{2}+\eta^2}
\Biggl\{2(k_t+p_t)^2\,e^{-is^\prime \frac{\psi-2\phi}{4}}\,
\sin\,\biggl(\frac{\psi-2\phi}{4}\biggr)\,\delta_{s^\prime\,r}\nonumber\\
& &+i(p_t-k_t)^2\Bigl(e^{-is^\prime \frac{\psi}{2}}+e^{-is^\prime \phi}\Bigr)
\,\delta_{s^\prime\,(-r)}\Biggr\}.
\end{eqnarray}
The result for $\omega(k\,r;p\,s)$ can be obtained
from  Eq. (\ref{B4}) after some replacements. A straightforward
calculation leads to

\begin{equation}
\label{B5}
\tilde{T}_2=i\frac{\pi\beta^2}{2m}
\delta_{ss^\prime}e^{-is\frac{\psi}{2}}
\int_0^{2\pi}d\phi\, \frac{\sin\bigl(\frac{\psi-2\phi}{4}\bigr)\;
\sin\bigl(\frac{\psi+2\phi}{4}\bigr)}
{\bigl[\sin^2\bigl(\frac{\psi-2\phi}{4}\bigr)+\nu^2\bigr]
\bigl[\sin^2\bigl(\frac{\psi+2\phi}{4}\bigr)+\nu^2\bigr]},
\end{equation}
where the integration over $k_t$ was performed via the theorem of 
residues, and  $\nu=\eta/2p_t$.
The remaining angular integral can be done as a contour
integration, after introducing the variable $x=e^{i\phi}$
(this is possible provided $\eta\rightarrow0$ only at the end):

\begin{eqnarray}
\label{B6}
& & \tilde{T}_2=i\frac{\pi^2\beta^2}{2m}
\delta_{ss^\prime}e^{-is\frac{\psi}{2}}
\frac{\sin\psi}{\sin\,\frac{\psi}{2}\,\bigl[
\cos\psi-1-8\nu^2-8\nu^4\bigr]}\,\frac{1}{\sqrt{1+\frac{1}{\nu^2}}},
\nonumber\\
& &\tilde{T}_2\, (\nu\rightarrow0)=0.
\end{eqnarray}
Eq. (\ref{B6}) agrees with the exact result 
for the scattering amplitude without radiative
corrections, which involves a factor of $\sin(\pi\beta)$
and thus excludes a contribution of order $\beta^2$ \cite{AB}.

To evaluate $T_1$,
we introduce form factors $F_1(q^2)$ and $F_2(q^2)$ for the
vertex function as
\begin{equation}
\label{5a}
\Gamma_\rho(q)= \gamma_\rho\, F_1(q^2)- \frac{1}{4 m} 
\,[\,\gamma_\rho,\gamma_\nu\,]\, q^\nu\, 
F_2(q^2)
\end{equation}
and after some standard manipulations one arrives at 

\begin{eqnarray}
\label{66}
& &-iT=-iT_1=-i(2\pi)^2\delta(q^0)\delta(q^3)\tilde{T},\\
& &\tilde{T}=-\pi\frac{\beta}{m}\Biggl\{
\biggl[1+F_1(q^2)-{\Pi}(q^2)\biggr]N(s,s^\prime)
+F_2(q^2)M_3(s,s^\prime)\Biggr\},\nonumber
\end{eqnarray}
where  \cite{Zuber} 
\begin{eqnarray}
\label{67}
& &F_1(q^2)=\frac{\alpha}{\pi}\Biggl[
\biggl(\ln\,\frac{\mu}{m}+1\biggr)\bigl(z\coth\,z-1\bigr)
-2\coth\,z\;\int_0^{\frac{z}{2}}d\xi\,\xi\tanh\,\xi-\frac{z}{4}
\tanh\,\frac{z}{2}\Biggr],\nonumber\\
& &F_2(q^2)=\frac{\alpha}{2\pi}\frac{z}{\sinh\,z},\\
& &{\Pi}(q^2)=-\frac{\alpha}{3\pi}\Biggl[\frac{1}{3}
+2\Biggl(1-\frac{1}{2\sinh^2\,\frac{z}{2}}\Biggr)\Bigl(\frac{z}{2}\coth\,
\frac{z}{2}-1\Bigr)\Biggr],\nonumber
\end{eqnarray}
with  $\vec{q}_t^{\phantom a 2}=4m^2\sinh^2\,\frac{z}{2}$ and 
$p\cdot p^\prime=m^2\cosh\,z$.
We have also defined the matrix elements

\begin{eqnarray}
\label{7}
& &M_3(s,s^\prime)=\bar{u}(p^\prime,s^\prime)\,\frac{i}{2}
[\gamma_1,\gamma_2]
\,u(p,s)\nonumber\\
& &\qquad\qquad =s\,\biggl(\,1-2e^{-is\frac{\psi}{2}}\cos\,\frac{\psi}{2}
\sin^2\,\frac{\theta}{2}\biggr)
\delta_{ss^\prime}-\frac{E}{m}\cos\,\frac{\psi}{2}\sin\,
\theta\;\delta_{s,-s^\prime},\nonumber\\
& &N(s,s^\prime)=i\epsilon^{ij}\,\frac{(p^\prime+p)_i\,q_j}
{\vec{q}_t^{\phantom
a 2}+\eta^2}\,
\bar{u}(p^\prime,s^\prime)\,u(p,s) +M_3(s,s^\prime)\nonumber\\
& &\qquad\quad\,\,\, =\frac{ie^{-is\frac{\psi}{2}}}{\sin\,
\frac{\psi}{2}}\,\delta_{s,s^\prime},
\end{eqnarray}
where we have used the helicity basis previously defined.

Using Eq. (\ref{7}),
the AB elastic scattering cross-section is given by

\begin{eqnarray}
\label{8}
& &\Biggl(\frac{d\sigma}{d\psi}\Biggr)_{\rm AB}
=\frac{m^2}{2\pi p}|\tilde{T}|^2\nonumber \\
& &=\Biggl(\frac{d\sigma}{d\psi}\Biggr)_{\rm AB}^{(1)}
\Biggl\{\,1+2\biggl[\,F_1(q^2)-{\Pi}(q^2)+F_2(q^2)\sin^2\,\frac{\psi}{2}\,
\biggr]\Biggr\}+O(\alpha^2),
\end{eqnarray}
where 
\begin{equation}
\label{9}
\Biggl(\frac{d\sigma}{d\psi}\Biggr)_{\rm AB}^{(1)}
=\frac{\pi\beta^2}{2p_t}\frac{1}{\sin^2\,\frac{\psi}{2}}\,\delta_{ss^\prime}
\end{equation}
is the  scattering cross-section in the  first order Born 
approximation \cite{Strings,Hagen,Vera}.

As it should, the scattering cross-section up to the order $\alpha$,  
Eq. (\ref{8}), displays helicity conservation. The first order Born 
approximation agrees with the weak flux limit of the exact
quantum-mechanical  AB scattering cross section 
of a spin 1/2 fermion given in \cite{Strings,Hagen,Vera},
in contrast with the results obtained for spinless particles 
\cite{Corinaldesi}. 

Since the electrical form factor $F_1$ becomes infrared divergent as
$\mu\rightarrow0$, it is necessary to include the inelastic
bremsstrahlung cross-section (see figures 1e and 1f) for soft photon production
in the (photon) energy range $\omega\leq \Delta E$,
where $\Delta E\,<<\,E$ is the typical energy resolution of
the detector. The calculations go along the same lines
as in the Coulomb scattering case \cite{Zuber}. Since we are only interested 
in the leading order terms in $\frac{\Delta E}{E}\,<<\,1$, whenever
allowed we shall take the limit $E^{\prime}=E-\omega\rightarrow E$ and
$|\vec{p}^{\phantom a \prime}|\rightarrow |\vec{p}|$.
With these approximations the scattering cross section factorizes in
the form ($v=\frac{p_t}{E}$):

\begin{eqnarray}
\label{10}
& &\Biggl[\frac{d\sigma}{d\psi}(\Delta E)\Biggr]_{\rm SB}
=\Biggl(\frac{d\sigma}{d\psi}\Biggr)_{\rm AB}^{(1)}\,
\frac{2\alpha}{\pi}\Biggl\{
\bigl(z\coth\,z-1\bigr)\ln\,\frac{2\Delta E}{\mu}
+\frac{1}{2v}\ln\,\frac{1+v}{1-v}\nonumber\\
& &\quad\quad
-\frac{1}{2}\cosh\,z\;\frac{1-v^2}{v\sin\,\frac{\psi}{2}}\,
\int_{\cos\,\frac{\psi}{2}}^1 d\xi\frac{1}
{(1-v^2\xi^2)\sqrt{\xi^2-\cos^2\frac{\psi}{2}}}
\ln\,\frac{1+v\xi}{1-v\xi}\Biggr\}.
\end{eqnarray}
The finite  inclusive AB cross-section
is defined by adding Eq.(\ref{8}) and Eq.(\ref{10}):

\begin{eqnarray}
\label{11}
& &\Biggl[\frac{d\sigma}{d\psi}(\Delta E)\Biggr]
=\Biggl(\frac{d\sigma}{d\psi}\Biggr)_{\rm AB}+
\Biggl[\frac{d\sigma}{d\psi}(\Delta E)\Biggr]_{\rm SB}\nonumber\\
& &\qquad\quad=\Biggl(\frac{d\sigma}{d\psi}\Biggr)_{\rm AB}^{(1)}
\Biggl(1-\delta_R\Biggr),
\end{eqnarray}
where $\delta_R$ is the contribution due to virtual plus
real (soft) photon emission, given by 

\begin{eqnarray}
\label{12}
& &\delta_R=\frac{2\alpha}{\pi}\Biggl\{
(1-z\coth\,z)\Bigl(1+\ln\,\frac{2\Delta E}{m}\Bigr)
-\frac{z}{2}\coth\,z\;\ln(1-v^2) -\frac{1}{9} \nonumber\\
& &+\frac{2}{3}
\Bigl(1-\frac{1}{2\sinh^2\,\frac{z}{2}}\Bigr)\biggl(1-\frac{z}{2}\coth\,
\frac{z}{2}\biggr)-\frac{1}{2}\frac{z}{\sinh\,z}
\sin^2\,\frac{\psi}{2}
-\frac{1}{2v}\ln\,\frac{1+v}{1-v}+\frac{z}{4}\tanh\,\frac{z}{2}\nonumber\\
& &+\frac{1-v^2}{2v}\frac{\cosh\,z}{\sin\,\frac{\psi}{2}}
\int_{\cos\,\frac{\psi}{2}}^1 d\xi\frac{1}
{\sqrt{\xi^2-\cos^2\frac{\psi}{2}}}
\biggl[ \frac{\ln(1+v\xi)}{1-v\xi}-\frac{\ln(1-v\xi)}{1+v\xi}\biggr]
\Biggr\}.
\end{eqnarray}
As shown in figure 2, we have plotted $\delta_R$ for three values of $E$. 
In that graph, contributions near $\psi=0$ have been omitted
since the AB amplitude is not well defined there \cite{forw}.

\section{Radiative Corrections to the Aharonov-Bohm Potential}
\label{SB}

In this section we shall compute vacuum polarization effects
(up to the order $\alpha$) on the renormalized
AB potential. In particular,
we will show that, as a result of radiative corrections,
a magnetic field is induced outside the flux tube.

Any external electromagnetic potential $A_\mu$ is modified due to
vacuum polarization effects. If this is properly taken  into account
the  effective AB  potential (${\cal A}_{\mu}$) becomes

\begin{equation}
\label{13}
{\cal A}_{\mu}(q)=(g_{\mu\lambda} 
- i \frac{\Pi_{\mu\lambda}}{q^2})\,A^\lambda(q)
=(1-\Pi(q^2))\,A_{\mu}(q)
\end{equation}

As the origin is a highly singular point in the
AB potential (\ref{2}) , it is convenient to introduce a
regularization which
distributes the magnetic flux on the surface of a 
thin cylindrical shell surrounding the origin,
similarly as done in \cite{Hagen}
($B_3(q)=(2\pi)^2\delta(q^0)\delta(q^3)\tilde{B}$),

\begin{eqnarray}
\label{14}
& &A^i(x_j)=-\frac{\Phi}{2\pi}\,\epsilon^{ij}\,\frac{x^j}{\rho^2}\,
\theta(\rho-R)\Longrightarrow
\tilde{A}^i(q_j)=i\Phi\,\epsilon^{ij}\,\frac{q^j}{q_t^2}\,J_0(R\,q_t),\\
\label{15}
& &B_3(x_j)=\frac{\Phi}{2\pi R}\,\delta(\rho -R)\Longrightarrow
\tilde{B}(q_j)=\Phi\,J_0(R\,q_t),
\end{eqnarray}
where $R$ is the radius of the cylindrical shell. As
$R\rightarrow0$ we recover the AB potential of Eq. (\ref{2}).

>From Eqs. (\ref{6}), (\ref{13}) and (\ref{14}) the potential 
induced by the vacuum polarization reads

\begin{equation}
\label{16}
A^{\rm vp}_i(x_j)=\Bigl(\frac{\Phi\alpha}{4\pi^3}\epsilon^{ij}\,
\partial_j\Bigr)\,\int_0^1 dz \frac{z^2}{1-z^2}
\,\biggl(1-\frac{1}{3}z^2\biggr)\,
\int d^2 q e^{i\vec{q}_t\cdot \vec{x}_t}\,\frac{1}{\frac{4m^2}{1-z^2}+q_t^2}
J_0(R\,q_t),
\end{equation}
with $A^{\rm vp}_0=A^{\rm vp}_3=0$. In Eq. (\ref{16}) we used the integral
representation \cite{Zuber}

\begin{equation}
\label{18}
\Pi(q^2)=\frac{\alpha}{\pi}\int_0^1 dz \frac{z^2}{1-z^2}
\,\biggl(1-\frac{1}{3}z^2\biggr)
\frac{q^2}{\frac{4m^2}{1-z^2}-q^2}.
\end{equation}
The integral over $d^2 q$ can be done in polar coordinates 
\cite{Grad}, with the result

\begin{eqnarray}
\label{19}
& &A^{\rm vp}_i(x_j)= -\frac{\Phi}{2\pi}\,\epsilon^{ij}\,\frac{x^j}{\rho^2}
\,\Biggl[\,\frac{2\alpha}{3\pi}\; G(m\rho)\,\Biggr],\\
& &G(m\rho)=m\rho\,\int_1^{\infty} \frac{dz}{z^3}\,\sqrt{2z^2+1}\,
I_0\bigl(2mRz\bigr)\,K_1\bigl(2m\rho z\bigr).\nonumber
\end{eqnarray}

The results may be translated in terms of an effective flux
$\Phi(\rho)$. In cylindrical coordinates, we obtain
from Eq. (\ref{19}) and Eq.(\ref{13}) the result

\begin{eqnarray}
\label{20}
& &{\cal A}_{\varphi}(\rho)=\frac{1}{2\pi\rho }\,\Phi(\rho),\nonumber\\
& &\Phi(\rho)=\Phi\,\Biggl[\theta(\rho-R)
+\frac{2\alpha}{3\pi}\,G(m\rho)\Biggr].
\end{eqnarray}
The integral in $G(m\rho)$ converges for $\rho>R\ne0$, and is 
logarithmic divergent
for $\rho=R$. A numerical solution is feasible, with the result
drawn in figure {\bf 3}. The effective flux  picture resembles
that of effective charge in QED. Indeed, from the figure 
we see that the vacuum fluctuations lead to a screening of the flux,
as a calculation of the induced magnetic field shows more clearly.
The total induced flux at large distance is given by the limit
$\rho\rightarrow\infty$ of the formula

\begin{equation}
\label{21}
\Delta\Phi(\rho)=\int_{C(\rho)}d\varphi\;\rho \, A^{\rm vp}_{\varphi}(\rho)
=\frac{2}{3}\frac{\Phi\alpha}{\pi}\,G(m\rho),
\end{equation}
and it is found to vanish due to the asymptotic behavior of $G(m\rho)$
for large distances.

As $R\rightarrow0$, it can be shown that $A^{\rm vp}_i(x_j)$ 
has an exact solution in terms of Meijer's $G$-function \cite{Grad},

\begin{equation}
\label{22}
A^{\rm vp}_i(x_j)= -\frac{\Phi}{2\pi}\epsilon^{ij}\frac{x^j}{\rho^2}
\Biggl[\frac{\alpha}{4\sqrt{\pi}}(m\rho)^5
\,G^{30}_{13}\Biggl( (m\rho)^2\Bigg| 
\begin{array}{rrr}
 & 0 &\\
-\frac{1}{2}&-\frac{1}{2}&-\frac{5}{2}
\end{array}\Biggr)\Biggr].
\end{equation}
The asymptotic limits for $m\rho<<1$ and $m\rho>>1$ can be calculated
from Eq. (\ref{22}) ($R=0$):

\begin{eqnarray}
\label{23}
& &\Phi(\rho)\approx \Phi\Bigl[1-\frac{2\alpha}{3\pi}\ln\,(m\rho)\Bigr]
\,\,\,\,(m\rho<<1),\\
& &\Phi(\rho)\approx \Phi\Bigl[1+\frac{\alpha}{4m\rho}\,e^{-2m\rho}\Bigr]
\,\,\,\,(m\rho>>1).
\end{eqnarray}

To calculate the induced magnetic field, we use $\vec{\cal{B}}=\nabla\times
\vec{\cal{A}}$ for $\rho\,>\,R$, and suppose that the general 
expression for $\vec{\cal{B}}={\cal B}_{3}\,\hat{z}$ is

\begin{eqnarray}
\label{24}
& &{\cal B}_{3}(\rho)=\frac{\Phi}{2\pi R}\delta(\rho-R)
+ B^{\rm vp}_3(\rho),\\
\label{25}
& &B^{\rm vp}_3 (\rho)=\Phi\gamma(R)\delta (\rho-R)
-\frac{2\Phi\alpha}{3\pi^2}m^2\,F(m\rho),\\
& &F(m\rho)=\int_1^{\infty} \frac{dz}{z^2}
\sqrt{z^2-1}(2z^2+1)\,I_0(2mRz)\,K_0(2m\rho z).\nonumber
\end{eqnarray}
The first term on the right hand-side of Eq.(\ref{24}) is
the original magnetic field in the cylindrical shell ($B_3$), whereas 
the vacuum polarization part $B^{\rm vp}_3$ contains
a highly concentrated contribution on the shell of the same relative
sign as $B_3$, together with an external contribution in the opposite
direction. In a semi-classical picture, this profile is expected since
the lines of the induced magnetic field are closed.
To determine the function 
$\gamma(R)$ we use that the total induced flux is zero. This gives
$\gamma(R)=\frac{\alpha}{3\pi^2 R}\, G(mR)$.

It is possible to use Eq.(\ref{22}) to deduce the asymptotic limits
for $B^{\rm vp}_3(\rho)$ in the case $R\rightarrow0$ and $\rho\neq0$
(at $\rho=0$ an additional divergent contribution has to be taken 
into account)

\begin{eqnarray}
\label{26}
& &B^{\rm vp}_3 (\rho)\approx -\frac{\Phi\alpha}{3\pi^2}\frac{1}{\rho^2}
\,\,\,\,(m\rho<<1),\\
& & B^{\rm vp}_3 (\rho)\approx
-\frac{\Phi\alpha}{4\pi}\frac{e^{-2m\rho}}
{\rho^2}\,\,\,\,(m\rho>>1).
\end{eqnarray}

\bigskip

\section{Conclusions}

We have calculated the order $\alpha$ radiative corrections, 
to the scattering cross section of an electron by an external 
AB potential, in first Born approximation.
The second order Born approximation was also calculated
and found to vanish, in agreement 
with the weak flux limit of the exact quantum mechanical
calculation (without radiative corrections).
The inclusive cross section, which besides the elastic scattering, 
also includes the soft photon bremsstrahlung was calculated. 
The \lq\lq semi-classical'',
radiatively corrected AB potential due vacuum polarization
currents and using a finite radius flux tube model was also calculated
up to the same order. It was shown that an induced magnetic field 
opposite to the original flux tube is induced outside, which partially
screens the original field.
We remark that our perturbative results are expected
to hold for $|\frac{e\Phi}{2\pi}|<<1$.

We would like to mention some literature which have contact with our work. 
Using the solutions of Dirac equation in the external AB potential, 
in reference \cite{Jasper} the processes of pair production and  
bremsstrahlung were analyzed. In \cite{Sere} the induced vacuum current was 
obtained. In these works the external AB potential was treated in an exact 
way, but radiative corrections due to virtual photons were not included. 
In our approach instead, the AB potential was perturbatively treated 
but radiative corrections were also included. 
It must be observed that our induced external magnetic field can 
alternatively be obtained from the induced vacuum current of \cite{Sere}, 
in the weak flux limit, through the use of the Ampere law. It must also 
be observed that the soft photon bremsstrahlung cross section, that is 
part of our inclusive cross section may be get from the bremsstrahlung 
cross section of \cite{Jasper} in the weak flux and
soft photon limits. Our radiative corrections due to virtual photons 
instead, namely the contributions of the graphs in figures 1b and 1c, are 
new and not included in these previous papers.

A more complete calculation of the AB cross section, in which radiative 
corrections are perturbatively considered as corrections to the exact 
treatment of the AB scattering is yet to be done.

\bigskip
\bigskip

LCA would like to thank the Mathematical Physics Department  
for their kind hospitality.  This work was partially supported by  
Funda\c c\~ao de Amparo a Pesquisa do Estado de S\~ao Paulo (FAPESP) and 
Conselho Nacional de Desenvolvimento Cient\'{\i}fico e Tecnol\'ogico (CNPq).

\newpage

\begin{center}

{\bf Figure captions}
\bigskip
\bigskip

\begin{itemize}

\item Figure 1. Diagrams contributing to:
(a) first order Born approximation , (b) vertex correction,
(c) vacuum polarization, (d) second order Born
approximation, and  (e)-(f) bremsstrahlung process.
The external field is represented by a cross.

\bigskip
\bigskip

\item Figure 2. $\delta_R$ for $m=0.5$ MeV and $\Delta E=0.1\; E$.

\bigskip
\bigskip

\item Figure 3. The function $G(m\rho)$,  
plotted for $m\,\rho\,>\,1$ and $m\,R=1$.

\end{itemize}

\end{center}


\begin{thebibliography}{100}

\bibitem{AB} Y. Aharonov and D. Bohm, Phys. Rev. {\bf 115}, 485, (1959);
M.Peshkin and A. Tonomura, {\it The Aharonov-Bohm
  Effect}, Lecture Notes in Physics {\bf 340} (Springer-Verlag, 1989).
\bibitem{Anyons}J.M. Leinaas and J. Myrheim, Nuovo Cimento {\bf B37}, 1, 
(1977); F. Wilczek, Phys. Rev. Lett. {\bf 48}, 1144, (1982); {\bf 49}
, 957, (1982); D.P. Arovas, R. Schrieffer, F. Wilczek and A. Zee,
Nucl. Phys. {\bf B251}, 117,  (1985). 
\bibitem{Strings} Ph. de Souza Gerbert, Phys. Rev. {\bf D40}, 1346, (1989); 
M. Alford and F. Wilczek, Phys. Rev. Lett. {\bf 62}, 1071, (1989).
\bibitem{Tono} A. Tonomura, Nuovo Cimento {\bf B110}, 571, (1995).
\bibitem{Jac} S. Deser, R. Jackiw and S. Templeton, Phys. Rev. Lett. {\bf 48}, 
975, (1982) and Ann. Phys. (N.Y.) 
{\bf 140}, 372, (1982); J.F. Schonfeld, Nucl. Phys. {\bf B185}, 157, (1981).
\bibitem{Lozano} O. Bergmann and G. Lozano, Ann. Phys. (N.Y.) 
{\bf 229}, 416, (1994).
\bibitem{Hagen0} C.R. Hagen, Phys. Rev. {\bf D56}, 2250, (1997).
\bibitem{MG1} M. Gomes, J.M.C. Malbouisson, A.J. da Silva,
Phys. Lett. {\bf A236}, 373, (1997), and Int. J. Mod. Phys. 
{\bf A13}, 3157, (1998).
\bibitem{Gomes1} M. Gomes and A. J. da Silva, Phys. Rev. {D57}, 3579, (1998);
H.O. Girotti, M. Gomes, J.R.S. Nascimento and A.J. da Silva,
Phys. Rev. {\bf D56}, 3623, (1997).
 \bibitem{Zuber} C. Itzykson and J.B. Zuber, {\it Quantum Field
Theory} (McGraw-Hill, 1985).
\bibitem{Hagen} C.R. Hagen, Phys. Rev. Lett. {\bf 64}, 503, (1990).
\bibitem{Vera} F. Vera and I. Schmidt, Phys. Rev. {\bf D42}, 3591, (1990); 
F.A. Coutinho and J. Fernando Peres, Phys. Rev. {\bf D49}, 2092,
(1994); H. Girotti and F. Fonseca Romero, Europhys. Lett. 
{\bf 37}, 165, (1997); see also Ph. de Souza Gerbert, ref. [3].
\bibitem{Corinaldesi} E.L. Feinberg, Sov. Phys. Usp. {\bf 5}, 753, (1963); 
E. Corinaldesi and F. Rafeli, Amer. J. Phys. {\bf 46}, 1185,
(1978); K. M. Purcell and W. C. Henneberger, 
  Amer. J. Phys. {\bf 46}, 1255, (1978).
\bibitem{forw} M.V. Berry et al. Eur. J. Phys. {\bf 1}, 154, (1980); 
R. Jackiw, Ann. Phys. {\bf  201}, 83, (1990);
D. Stelitano, Phys. Rev. {\bf D51}, 5876, (1995); P. Giacconi,
F. Maltoni and R. Soldati,  Phys. Rev. {\bf D53}, 952, (1996);
S. Sakoda and M. Omote, J. Math. Phys. {\bf 38}, 711, (1997).
\bibitem{Grad} I.S. Gradshteyn and I.M. Ryzhik, 
{\it Table of Integrals, Series, and Products}
(Academic Press Inc., Fifth Edition, 1994). 
\bibitem{Jasper} J. Audretsch, U. Jasper, and V.D. Skarzhinsky,
Phys. Rev. {\bf D53}, 2178, (1996) ; {\bf Ibid.} 2190.
\bibitem{Sere} E.M. Serebryanyi, Theor. Math. Phys. {\bf 64}, 846, (1986).

\end{thebibliography}
\end{document}